# Optimal Placement of Valves in a Water Distribution Network with CLP(FD)

MASSIMILIANO CATTAFI, MARCO GAVANELLI,

MADDALENA NONATO, STEFANO ALVISI and MARCO FRANCHINI

*Department of Engineering*
*University of Ferrara*
*Via Saragat, 1*
*44122, Ferrara, Italy*

## Abstract

This paper presents a new application of logic programming to a real-life problem in hydraulic engineering. The work is developed as a collaboration of computer scientists and hydraulic engineers, and applies Constraint Logic Programming to solve a hard combinatorial problem. This application deals with one aspect of the design of a water distribution network, i.e., the valve isolation system design.

We take the formulation of the problem by Giustolisi and Savić (2008) and show how, thanks to constraint propagation, we can get better solutions than the best solution known in the literature for the Apulian distribution network.

We believe that the area of the so-called *hydroinformatics* can benefit from the techniques developed in Constraint Logic Programming and possibly from other areas of logic programming, such as Answer Set Programming.

*KEYWORDS*: Constraint Logic Programming, hydraulic engineering, valve placement, graph partitioning.

## 1 Introduction

An aqueduct is a complex system that includes a main component to transport water and a water distribution component, that brings the water to the users. The water distribution network can be thought of as a labelled graph, in which pipes are represented as undirected edges. In the network, there is at least one special node that is the source of the water (node 0 in Figure 1); users are then connected to the edges of the water distribution network. Each user has a demand (in litres per seconds) that is quantified by the hydraulic engineer through the available data. In particular, such a demand is frequently expressed as a daily average value. Each edge of the graph is labelled with the total demand of the users linked to it. For example, in Figure 1, the edge connecting nodes 2 and 5 (let us name it $e_{2,5}$) has a demand of $15l/s$ (that may be due, e.g., to five clients each requesting $3l/s$ on average).

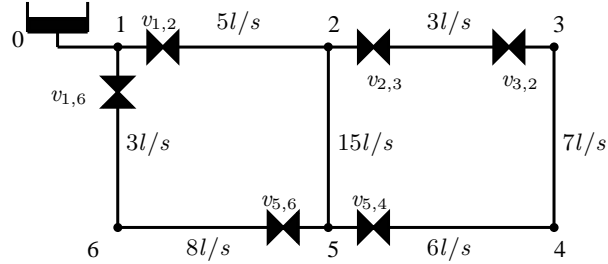

Fig. 1. A schematic water distribution system with valves

When designing a water distribution network, one of the steps is designing the isolation system: in case a pipe has to be repaired (e.g., because of a break), a part of the network has to be disconnected from the rest of the network, in order to allow workers to fix the broken pipe. The isolation system consists of a set of *isolation valves*, that are placed in the pipes of the network. Once closed, the isolation valve blocks the flow of water through the valve itself. In common practice, a valve is usually placed in a pipe near one of the two endpoints; this means that in each pipe at most two valves can be placed. If in some pipe there are two valves, this means that this single pipe can be isolated by closing both the valves. In the example of Figure 1, the edge $e_{2,3}$ connecting nodes 2 and 3 has two valves, so in case this pipe is damaged, valves $v_{2,3}$ and $v_{3,2}$ will be closed, isolating only $e_{2,3}$.

However, placing two valves in each pipe is often not a viable option, because each valve has a cost; the cost is not only due to the manufacturing and physical placing of the valve, but also to the fact that the pipe is more fragile and deteriorates more quickly near valves. In case there are not two valves in each pipe (as it is usually the case in real distribution networks), the isolation of a pipe implies the closure of more than two valves and thus the isolation of more than one pipe. In this case, more users other than those connected to the broken pipe will remain without service during pipe substitution. Suppose that the pipe $e_{3,4}$ connecting nodes 3 and 4 is damaged. In order to de-water it, workers have to close valves $v_{3,2}$ and $v_{5,4}$; as a result edge $e_{5,4}$ will be de-watered as well, and the clients that take water from it will have no service as well. Valves partition the network in the so-called *sectors*, that are, intuitively, those parts of the distribution network enclosed by some set of valves: edges $e_{3,4}$ and $e_{4,5}$ are in the same sector, so they cannot be de-watered independently one from the other.

The usual measure of the disruption in the service is the *undelivered demand*, i.e., the demand (in litres per second) that is not fulfilled during the repair operations; in the case there is need to de-water edge $e_{3,4}$, the disruption is the demand of the edges $e_{3,4}$ and $e_{4,5}$, i.e., $7 + 6 = 13l/s$. However, notice that the undelivered demand does not always coincide with the sector the damaged pipe belongs to. For example, pipe $e_{2,5}$ belongs to the sector consisting of the edges $e_{1,2}$ and $e_{2,5}$, that is surrounded by valves $v_{1,2}$, $v_{2,3}$, $v_{5,4}$, and $v_{5,6}$; however by closing these four valves, we will de-water a larger part of the network: edges $e_{2,3}$, $e_{3,4}$, and $e_{4,5}$ will be de-watered even though they are not in the same sector of the broken pipe. This effect is called *unintended isolation* (Jun and Loganathan 2007).

The design of the isolation system consists of placing in the distribution network a given number of valves such that, in case of damage, the disruption is "minimal". Of course, the level of disruption depends on which pipe has to be fixed. In Figure 1 we have four sectors: if $e_{2,3}$ is damaged, the undelivered demand during repair is $3l/s$, if one of $\{e_{3,4}, e_{4,5}\}$ is broken, the undelivered demand is $13l/s$, if the broken pipe is $e_{1,6}$ or $e_{5,6}$ the undelivered demand is $3 + 8 = 11l/s$, while for sector $\{e_{1,2}, e_{2,5}\}$ the undelivered demand is $36l/s$, corresponding to the demand of $\{e_{1,2}, e_{2,5}, e_{2,3}, e_{3,4}, e_{4,5}\}$. A usual measure (Giustolisi and Savić 2010) is to take the worst case, and assign to the placement shown in Figure 1 (characterised by 6 valves) the effect of the maximal possible disruption: $36l/s$.

Giustolisi and Savić (2010) address the design of an isolation valve system as a two-objective problem: one objective is minimizing the number of valves in the isolation system, and the other is the minimization of the (maximum) undelivered demand. They adopt a genetic algorithm that is able to provide near-Pareto-optimal solutions, and apply it to the Apulian distribution network. The genetic algorithm provides good solutions in a very short time, but it is incomplete, so it does not provide, in general, Pareto-optimal solutions, but only solutions that are hopefully near to the Pareto front. The real optimal Pareto front remains unknown.

We believe that a complete search algorithm could provide better solutions, although at the cost of a higher computation time. Since the problem should be solved during the design of the valve system, there is no need to have a solution in real-time, and an algorithm providing a provably Pareto-optimal solution may be preferable with respect to incomplete algorithms, even with higher computation times.

In this paper, we address the same two-objective problem studied by Giustolisi and Savić (2010) as a sequence of single-objective ones; this is always possible when one of the objectives is integer (Van Wassenhove and Gelders 1980; Gervet et al. 1999; Gavanelli 2002). Given the number of valves, we model the design of the isolation valve system as a two-player game, and solve it with a minimax approach (Russell and Norvig 2003). As the game has an exponential number of moves, we reduce the search space by pruning redundant branches of the search tree, implementing the minimax algorithm in Constraint Logic Programming (CLP) (Jaffar and Maher 1994) on Finite Domains (CLP(FD)) (Marriot and Stuckey 1998; Frühwirth and Abdennadher 2003; Dechter 2003), in particular we used ECL$^i$PS$^e$ (Apt and Wallace 2006). Our algorithm is complete, so it is able to find the optimal solutions and prove optimality; we show improvements on the best solutions known in the literature, up to 10% of the objective function value.

The rest of the paper is organized as follows. In the next section, we give a formal description of the problem, then we propose the minimax interpretation in Section 3. We give a CLP(FD) model in Section 4, then we detail some implementation issues in Section 5. Section 6 is devoted to experimental results. We discuss related work in Section 7 and then we conclude.

## 2 Problem description

A water distribution network is modelled as a weighted undirected graph $G \equiv (N, E)$, where $N = \{1, \ldots, n\}$ is a set of nodes and $E = \{e_{ij}\}$ is a set of edges. Each edge $e_{ij}$ has an associated weight $w(e_{ij})$ called *demand*.

In the network, there are some nodes identified by the set $\Sigma$ that are called *sources*.

Valves can be positioned near one of the ends of a pipe; we will refer to valve on edge $e_{ij}$ near to node $i$ as $v_{ij}$, while $v_{ji}$ is a valve on the same edge, but close to node $j$.

Given a number $N_v$ of valves to be positioned in the network, the objective is to position the valves in the network such that:

1. it is possible to isolate any pipe in the network. Formally, given an edge $e_{ij}$, it is possible to identify a set of valves $C$ to be closed such that there is no path from any source node $s \in \Sigma$ to the edge $e_{ij}$ that does not contain a valve $v \in C$. Since the set $C$ of valves to be closed depends on the damaged pipe $e_{ij}$, we will also write $C(e_{ij})$. Note that there is only one reasonable set $C(e_{ij})$ of valves to be closed given a broken edge $e_{ij}$: intuitively only the valves directly reachable from $e_{ij}$ will be closed. For example, in Figure 1 if the broken edge is $e_{3,4}$ then $C(e_{3,4}) = \{v_{3,2}, v_{5,4}\}$ and it does not make sense to close farther valves, such as $v_{2,3}$, because in order to reach $v_{2,3}$ from $e_{3,4}$ we have to overpass other valves ($v_{3,2}$).

2. the objective is to minimize the maximum undelivered demand (UD). Formally, let $D(C)$ be the set of edges that do not receive water when the valves in $C$ are closed, i.e., those edges for which there is no path from any source node to the edge: $D(C) = \{e_{ij} \in E | \forall s \in \Sigma, \nexists Path(s, e_{ij})\}$. The objective function to be minimized is

$$UD = max_{e_{ij} \in E} \sum_{e_{kl} \in D(C(e_{ij}))} w(e_{kl}).$$

## 3 Game model

The problem can be considered as a two-player game, consisting of the following three moves:

- the first player decides a placement of $N_v$ valves in the network;
- the second player selects one pipe to be damaged;
- the first player closes a set of valves that de-waters the damaged pipe.

The cost for the first player (and reward for the second) is the undelivered demand: the total demand (in litres per seconds) of all users that remain without service when the broken pipe is de-watered.

Given this formalization, the well-known minimax algorithm is applicable (Russell and Norvig 2003).

As we said in Section 2, choosing the last move is very easy, as there is only one reasonable solution: close all valves that are reachable from the broken pipe, without overpassing other valves. An implementation of this algorithm is given by Jun and Loganathan (2007).

Clearly the first step of the first player is the most sensitive, because it can generate a wide number of alternatives. In a network with $N_e$ edges and with $N_v$ valves, the search space is $\binom{2N_e}{N_v}$, since each edge can host up to two valves. However, some of the moves are not very interesting, for three main reasons, that will be explained in detail in the next section. First, some solutions are clearly non-optimal. Second, some are symmetric, and provide valve placements that, although different, represent equivalent solutions. Third, after some solution is known, there is no point in looking for worse solutions: as soon as the current search branch cannot lead to solutions better than the incumbent, we can stop the search, backtrack, and continue from a more promising branch.

Each of these three cases provides a possible pruning of the search space, that can exponentially speed-up the computation with respect to a naive approach. The first two cases can be thought of as *constraints*, while the third can be though of as a *bound*: all of them can be simply cast in Constraint Logic Programming on Finite Domains (CLP(FD)).

## 4 Constraint Logic Programming model

We can now show how to model in CLP(FD) the valve placement problem. We first provide a simple minimax algorithm, then improve it with the three types of pruning hinted at earlier.

### *4.1 A minimax implementation in CLP(FD)*

We associate a Boolean variable to each possible position of a valve (so we have two Boolean variables for each edge in the graph); if the variable takes value 1, then the given end of the edge hosts a valve, otherwise, if the variable takes value 0, there is no valve in such location. In the following, the list of these variables is called *Valves*.

The two-player game can be implemented as follows:

```
solve(Valves,N_v):-
   impose_constraints(Valves,N_v),
   minimize(
      (  assign_valves(Valves),
         maximize(
            (  break_pipe(Broken),
               close_valves(Valves,Broken,ClosedValves),
               undelivered_demand(Valves,ClosedValves,UD)
            ), UD, MaxUD)
      ),MaxUD,MinMaxUD).
```

The `minimize/3` and `maximize/3` meta-predicates are predefined in most CLP(FD) languages; declaratively, `minimize(G,F,V)` provides, amongst the solutions of goal $G$ (bindings to the variables in $G$ that make true the goal $G$), the solution that provides the minimum value for variable $F$ (Marriott and Stuckey 1994; Fages 1996); such minimal value is bound to variable $V$. It is equivalent to the ASP syntax $\sharp min(F : G) = V$ (Faber et al. 2008). In other words, the result of $minimize(G, F, V)$ is equivalent to the Prolog goal

$$findall(F, G, List), \, minlist(List, V)$$

where *minlist* finds the minimum value $V$ in the *List*.

Operationally, it has a better performance, since it does not need to find all the solutions of $G$, collect them in a *List*, and find the minimum, but it implements a form of branch-and-bound. Operationally, `minimize` calls goal $G$ and, if it succeeds providing some binding $F/F^*$, it imposes a new unbacktrackable constraint $F < F^*$; then it continues the search. The unbacktrackable constraint is considered in the constraint store of all the nodes of the search tree, and prunes every node that cannot possibly provide a lower value than $F^*$. When the goal $G$ fails, the optimal value is the last value obtained as $F^*$ (Prestwich 1996), if it exists. `maximize` is treated symmetrically.

Predicate `impose_constraints` posts all the constraints of the model to the constraint solver. It contains the constraint stating that there are $N_v$ valves in the distribution network; other constraints will be described in Section 4.2.

`assign_valves` starts the search on the *Valves* variables.

After finding a possible positioning of the valves that satisfies all constraints, a maximisation phase tries the moves of the opponent player: it searches (predicate `break_pipe`) the pipe that, if damaged, can be fixed only giving a maximum disruption of the service. When the opponent has chosen a pipe to break, we can compute the valves that should be closed to allow for substitution of the *Broken* pipe; finally, we compute the undelivered demand.

Thus, the internal `maximize` finds, amongst the moves of the opponent player (`break_pipe`), the move that gives the maximum undelivered demand; such value is bound to variable *MaxUD*. The first player, instead, chooses the placement of the valves (`assign_valves`) with the aim of minimizing the value *MaxUD*.

### *4.2 Reducing the number of moves*

The number of moves of the first player is huge even for small networks and number of valves. However, as hinted at earlier, some configurations can be avoided, as shown in the next paragraphs.

#### *4.2.1 Redundant valves and symmetries*

Consider the network in Figure 2. Even without knowing the demand on the various pipes, we can tell that some of the valves are redundant, just by looking at the topology of the network.

Valve $v_{1,2}$ cannot be used to identify a sector: the pipe immediately on the left of the valve belongs to the same sector as the pipe immediately on the right. In fact, there is a closed path going from one side of the valve to opposite side: starting from node 1, we can go to node 3, then 4, then 2, and we reach the opposite side of the same valve without having met any other valve. The same holds for valve $v_{3,5}$: there exists the path $(3, 4, 6, 5)$ that connects one end of the valve to the other end.

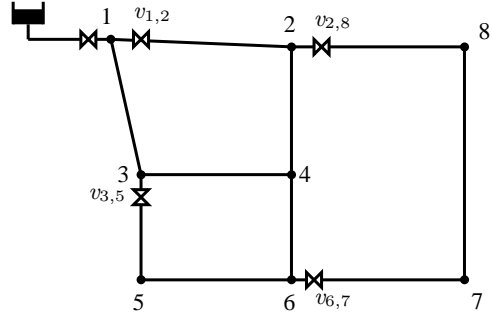

Fig. 2. A network with redundant valves

In general, we can say that in any closed path of the network, there cannot be exactly one valve. No valves means that the whole path will be contained in a sector, which is sensible. Two valves or more can mean that the path is divided into two or more sectors. So, for each closed path, one could impose a constraint saying that the number of valves in such path cannot be equal to 1.

Indeed, the number of paths is exponential in the size of the network, however we can choose to impose such constraint only for a limited number of closed paths. We decided to impose one such constraint for each (boundary of a) *face*, that is a concept of planar graphs. When drawing the graph on a plane, each of the regions surrounded by edges of the graph is called a *face*. The number of faces of a planar graph is always polynomial, as proven by Euler.

Notice that, when a node is connected to exactly two edges, we have a symmetry. For example, consider node 8 in Figure 2: in one assignment, we could have a valve $v_{8,2}$, while another assignment could be identical but with a valve in $v_{8,7}$. These two solutions are symmetric, because the fact that node 8 is in the same sector as edge $e_{2,8}$ or as $e_{7,8}$ is irrelevant, since nodes do not have a contribution to the objective function. So, we can impose the symmetry breaking constraint $v_{8,7} = 0$. This simple observation can provide a notable speedup in the search, because real networks often have this situation.

### *4.2.2 Bounding*

Consider a node in the search tree that selects the move for the first player (predicate `assign_valves/1`): in a generic node, some of the $v_{ij}$ variables will be assigned value 1 (meaning that some valves have already been placed), some variables will have value 0 (meaning that in such position there is no valve), and some will still be unassigned.

Consider the example in Figure 3: circles represent positions in which there is no valve, while strokes are variables still unassigned. Even though we do not have a complete placement, we can already say that there is a sector containing at least edges $e_{7,8}$, and $e_{6,7}$. The opponent player will have the option of damaging, e.g., pipe $e_{7,8}$, causing an undelivered demand that is no less than $w(e_{7,8}) + w(e_{6,7})$. So if the cost of such sector is worse than the current best solution found by the first player (i.e., $w(e_{7,8}) + w(e_{6,7}) > UD^{best}$), there is no point in continuing the search on the current branch. Note that this bound considers only the cost of the sector, without including unintended isolation.

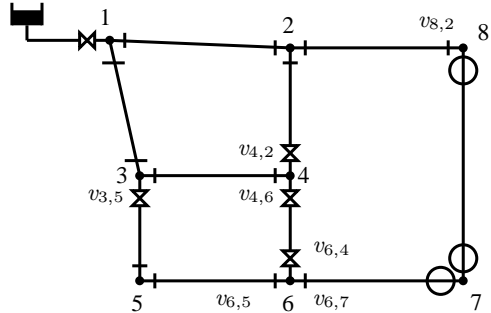

Fig. 3. A partial assignment: circles mean absence of valve, strokes are variables not assigned yet

We can also perform a reasoning similar to reduced costs pruning (Focacci et al. 1999; Focacci et al. 2002). Suppose that $w(e_{7,8}) + w(e_{6,7}) < UD^{best}$ but adding $w(e_{2,8})$ is enough to overpass the current best $UD^{best}$ (i.e., $w(e_{2,8}) + w(e_{7,8}) + w(e_{6,7}) > UD^{best}$): this means that we cannot afford to include edge $e_{2,8}$ in the same sector, and the only possibility to get a solution better than $UD^{best}$ is to separate the two sectors, placing a valve in $v_{8,2}$. Thus, we can impose $v_{8,2} = 1$.

# 5 Implementation details

## 5.1 Incremental bound computation

The bound described in Section 4.2.2 is very powerful, and reduces significantly the number of explored nodes. However, it can be rather expensive in terms of computing time, if implemented naively. In fact, it implies computing the cost of the sector one edge belongs to, which means identifying the sector, possibly visiting a significant part of the graph. So computing it again and again during search can make it very time consuming.

Instead of restarting from scratch the identification of the sectors and computing their cost at each node of the search tree, we compute them incrementally.

We associate to each node $i$ of the graph a variable $S_i$ that represents the sector the node belongs to, and the lower bound $LB_i$ on the cost of the sector $S_i$.

A constraint is associated with each edge of the graph $e_{i,j}$, and relates the two variables $v_{i,j}$ and $v_{j,i}$ with the two sectors $S_i$ and $S_j$, and with their lower bounds $LB_i$ and $LB_j$:

$$lower\_bound(v_{ij}, v_{ji}, S_i, S_j, LB_i, LB_j) \cdot \tag{1}$$

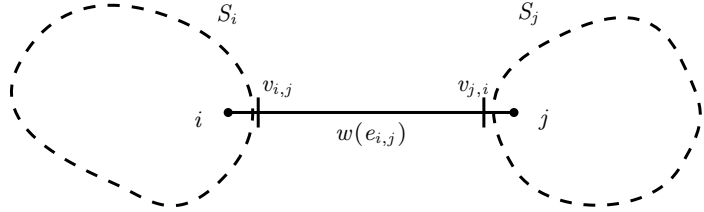

Fig. 4. Example of propagation of the lower bound when joining sectors

Declaratively, constraint (1) states that, given the value of variables $v_{i,j}$ and $v_{j,i}$, the undelivered demand cannot be lower than (the maximum of) the two bounds $LB_i$ and $LB_j$.

Operationally, the constraint (1) is awakened when one of the variables $v_{i,j}$ or $v_{j,i}$ becomes ground.

Initially no valve is placed, and each node is (tentatively) a sector by itself with associated lower bound zero (since no demand is associated to nodes).

If variable $v_{i,j}$ takes value 0, this means that there will be no valve in such position, so the sector $S_i$ will have to include edge $e_{i,j}$, and we increment the value of the lower bound $LB_i$ by $w(e_{i,j})$ (see Figure 4).

If both variables $v_{i,j}$ and $v_{j,i}$ have value 0, this means that the two sectors $S_i$ and $S_j$ should be joined: we unify the corresponding variables $S_i = S_j$, and increment the value of the lower bounds: we compute the cost of the joined sector as $LB_i + LB_j + w(e_{i,j})$.

Moreover, as explained in Section 4.2.2, if $v_{i,j} = 0$ and $v_{j,i}$ is not ground yet, but $LB_i + LB_j + w(e_{i,j})$ is greater than the current best solution, then joining the two sectors would give a solution worse than the current best, so we can impose a valve near node $j$, i.e., $v_{j,i} = 1$.

### *5.2 Dealing with unintended isolation*

As mentioned in Section 4.2.2, the bound computed by constraint (1) does not take into account unintended isolation. However, when evaluating the total damage associated to the breaking of a certain pipe (which results in the isolation of a certain sector) it is necessary to consider also this aspect. Predicate `undelivered_demand/3` finds the correspondent actual value of the objective function used in `maximize/3`.

Its algorithm is based on the following principle. Isolating a sector is equivalent to removing, from the graph describing the network, the part of the graph which belongs to said sector. It is then subsequently possible to determine the connected components of the obtained subgraph (computational complexity is linear in the size of the subgraph (Hopcroft and Tarjan 1973)). The `graph_algorithms` library (ECL$^i$PS$^e$ documentation ) of ECL$^i$PS$^e$ Prolog provides efficient implementations of predicates for such operations on graphs.

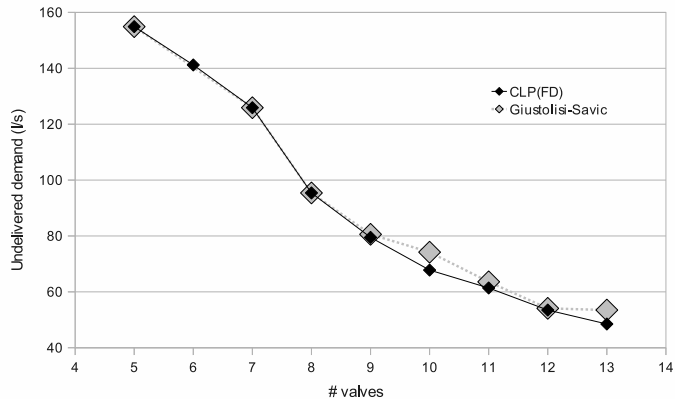

Fig. 5. Comparison between the approximate Pareto front computed by Giustolisi-Savić and the optimal Pareto front obtained in CLP(FD)

Selecting the connected component which includes the source node and summing up the demands on the pipes contained in it gives the *deliverable* demand. The total undeliverable demand can thus be obtained subtracting the deliverable demand from the total network demand.

## 6  Experimental results

We compare our results with those reported by Giustolisi and Savić (2008), and we apply our CLP(FD) algorithm on the Apulian water distribution network reported in that paper. Both the software and the instance are available at the web page (Cattafi and Gavanelli 2011). The network has 23 nodes and 33 edges. Giustolisi and Savić (2008) adopt a multi-objective genetic algorithm, that minimizes both the number of valves and the undelivered demand. The aim is to find the so-called Pareto frontier (Gavanelli 2002); in this problem, a solution belongs to the frontier if there is no way to reduce the undelivered demand without increasing the number of valves (and, vice-versa, it is impossible to reduce the number of valves without increasing the undelivered demand). The genetic algorithm, however, is not able to prove that a solution is indeed Pareto-optimal, and provides an approximation of the Pareto frontier, i.e., a set of points that are hopefully near to the real Pareto frontier. Moreover, Giustolisi and Savić (2008) use a simplifying assumption: *"in order to reduce greatly the search space of the optimizer, the constraint of a maximum of one valve for each pipe was tested"*. In the paper, they report the best found solutions obtained with a number of valves ranging from 5 to 13.

We computed the true Pareto-optimal frontier by varying the number of valves from 5 to 13 valves, and computing for each value the best placement. The comparison of the near-Pareto-optimal frontier and the true Pareto-optimal frontier obtained with our CLP(FD) program is shown in Figure 5. It is worth noting that Giustolisi and Savić (2008) do not provide a solution with 6 valves, possibly because their algorithm was not able to find a solution with undelivered demand lower than that obtained with 5 valves. We proved, instead, that such a solution exists and adding a valve reduces the damage. Excluding this case, when the number of valves is low (up to 8 valves) their algorithm found the real optimum, probably due to the fact that the search space is still not very wide, so the genetic algorithm is able to explore a wider percentage of the search space. When the number of valves increases, their algorithm gets farther from the real optimum, with a gap of about 10% with 10 and 13 valves. Note also that we were able to find a solution with 12 valves that gives the same undelivered demand that Giustolisi and Savić (2008) compute with 13 valves: in this sense, we were able to save one valve (out of 13) maintaining the same cost for undelivered demand.

The computation time is reported in Figures 6 (linear scale) and 7 (log scale). All experiments were done on a computer featuring an Intel Core 2 Duo T7250 2GHz processor with 4GB of RAM (note, however, that the current implementation does not use parallelism, and uses only one core). We show the performance of the basic algorithm, and of the improved versions that include the reduction of redundant valves (Section 4.2.1) and the bound (Section 4.2.2) varying the number of valves. The graphs also show the performance of another implementation of branch-and-bound available in ECL$^i$PS$^e$, called *min_max*. From the graph in linear scale (Figure 6) we can see that each of the improvements has a significant impact in terms of reduction of the computation time. When the number of valves is low, the elimination of redundant valves (i.e., imposing that in a face there cannot be exactly one valve, Section 4.2.1) has a very strong effect, while the bound has almost no effect. On the other hand, when the number of valves increases, the bound seems to have a higher impact. Combining the two, we get a further improvement, with a reduction of the computation time of more than two orders of magnitude.

Figure 7 shows that the computing time grows less than exponentially with respect to the number of valves. This can be explained by the fact that the search space does not grow exponentially, but it varies as the binomial coefficient.

ECL$^i$PS$^e$ has two implementations of the branch-and-bound predicate for minimization (Prestwich 1996). One, called `min_max`, restarts the search after a new solution is found; this means that the first part of the search tree is explored every time a new solution is found; on the other hand, restarting the search allows ECL$^i$PS$^e$ to add the unbacktrackable constraint from the root node of the search tree, and propagate effectively on all the nodes of the tree. The second, called `minimize`, avoids the restarts and continues the search, taking the risk that the newly added unbacktrackable constraint will not be able to propagate immediately, but only after some changes to the domains of the cost variable has happened. In our application, we found that `min_max` was about one order of magnitude slower than `minimize`, as shown in Figures 6 and 7.

Indeed the computation time is much higher than that reported by Giustolisi and Savić (2008): they computed the whole (near) Pareto frontier in just 10 minutes, on an older computer. However, our algorithm is able to find the true optimum and prove its optimality, which is well-known to be often more difficult than finding the optimum itself, so in a fair comparison the time required for proving optimality should not be taken into account. In Figure 8 we show the anytime behaviour of our algorithm, i.e., we plot the solution quality with respect to the computing time in a typical instance. Indeed, our algorithm takes about 50 minutes to compute just one point of the Pareto frontier. However, looking closer at the graph one notices that our algorithm gets to a reasonable quality in a few seconds, it takes 27 minutes to get to the same quality obtained by the genetic algorithm, then it is able to improve on it and takes 37 minutes (total) to find the real optimal solution.

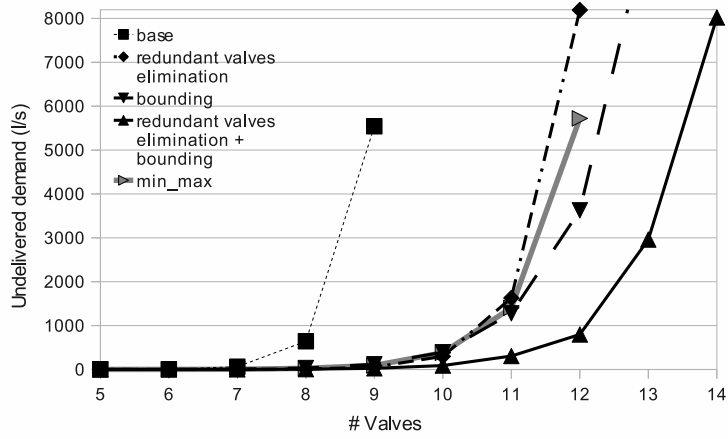

Fig. 6. Computation time of the algorithms including different optimizations

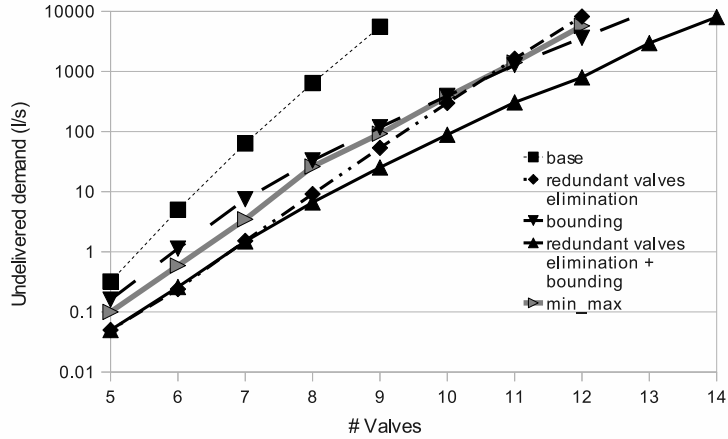

Fig. 7. Computation time of the algorithms including different optimizations, log scale

Giustolisi and Savić (2008) also show the graph of the (near) Pareto-optimal frontier with a higher number of valves, but they do not report the solutions, so we cannot make a comparison. We tested our algorithm with the same number of valves (up to 24); we could not prove optimality, but we were able to find reasonable solutions within a few minutes.

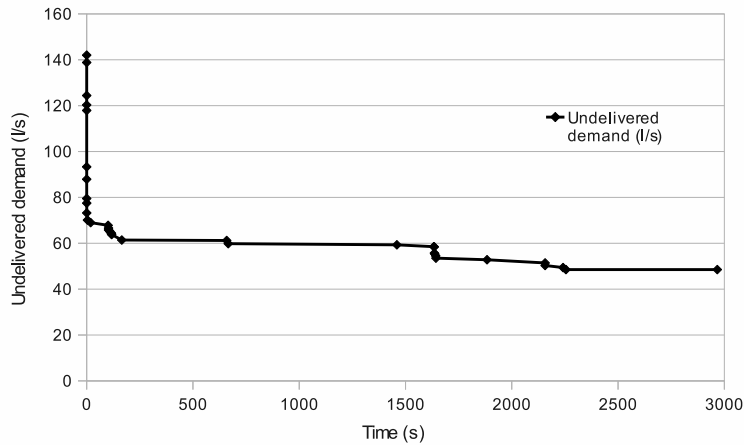

Fig. 8. Anytime behaviour of the CLP(FD) algorithm: solution quality with respect to the computation time. Number of valves $N_v = 13$

## 7 Related work

In the literature of hydraulic engineering, two main problems related to the isolation valves in a pipe network have been faced, that is a) the identification of the segments and undesired disconnections that occur after a set of isolation valves has been closed and b) the (near) optimal location of the set of isolation valves.

As far as the first topic concerns, in the literature there are a number of studies regarding segment identification and the undesired disconnections that occur following the closure of a set of isolation valves. In particular, the methods proposed by Jun and Loganathan (2007) and Kao and Li (2007) are based on a dual representation of the network, with segments treated as nodes and valves as links. The methods proposed by Creaco et al. (2010) and Giustolisi and Savić (2010) use topological incidence matrices to identify the segments.

As far as the second topic concerns, recently, Giustolisi and Savić (2010) have presented a method for the near-optimal placement of isolation valves based on a multi-objective genetic algorithm. Given that the placement of isolation valves is the result of a compromise between the need to reduce the costs tied to purchasing and installing the valves and the simultaneous need to assure high system reliability in the event of routine or non-routine maintenance, Giustolisi and Savić (2010) use the number of valves to be installed–as a surrogate for cost– and the maximum demand shortfall (the demand shortfall represents the unsupplied water demand after isolating a segment) in the different (disconnected) segments of the network as the objective functions to be minimised. Creaco et al. (2010) instead propose a different couple of objective functions, that is total cost of the set of valves, the cost of each valve being calculated as a function of the pipe diameter, and the weighted average unsupplied demand associated with the segments. Also in (Creaco et al. 2010) the optimization is solved through a multi-objective genetic algorithm. All of these works use incomplete algorithms, that cannot ensure that the found solution is the real optimum; to the best of our knowledge, this work presents the first complete algorithm to address the valve placement problem.

The valve placement problem has some similarities with the graph partitioning problem, in which the goal is to partition a graph into (almost) equal-size parts by removing the minimal number of edges or (in the weighted edges case) such that the total weight of the edges which connect different parts is minimized. In general, graph partitioning is NP-hard (Garey and Johnson 1990). Most works in the literature deal with heuristics or approximation algorithms.

One of the first works in the area is by Kernigan and Lin (1970), that propose a greedy algorithm which outputs a graph bisection. Starting from an initial solution (which can be suggested by some criterion or also be found randomly), each step of the algorithm evaluates the improvement in the objective function that would be obtained moving a vertex from one partition to the other and takes the best choice. Iteration goes on until convergence to a local optimum is reached. The bisection can be applied recursively to partition further. Fiduccia and Mattheyses (1982) improve the algorithm so that the asymptotic behaviour of the algorithm is linear rather than quadratic.

A different approach is based on the *spectral* analysis of the graph. A graph can be represented by its incidence matrix: a square matrix $N \times N$ (if $N$ is the number of vertices) whose $(i, j)$ element is 1 if there is an edge from vertex $i$ to vertex $j$ and 0 otherwise. Its representation as a Laplacian matrix is obtained as the (matricial) difference between the diagonal matrix which has in position $(i, i)$ the degree of the node $i$, and the incidence matrix. The set of the eigenvalues of the Laplacian is the graph *spectrum* (Chung (1994) gives an extended tractation of the subject). Since it was shown (Fiedler 1973) that the second smallest eigenvalue of the Laplacian associated to a graph contains information about its connectivity, various partitioning heuristics were proposed relying on the eigenvectors (Hendrickson and Leland 1995a; Spielman and Teng 1996). In comparison with other heuristics, spectral methods provide good quality partitions at an increased computational cost (necessary to compute the matrix eigenvalues).

Various kinds of heuristics can be used in multilevel schemes, which reduce the size of the graph by collapsing vertices and edges, partition the smaller graph, and then uncoarsen it to construct a partition for the original graph. Hendrickson and Leland (1995b) employ spectral methods to partition the smaller graph, and use a variant of the Kernighan-Lin algorithm to periodically refine the partitions. Karypis and Kumar (1998) adopt a coarsening heuristic for which the size of the partition of the coarse graph is within a small factor of the size of the final partition.

The special case of *planar* graphs (i.e. graphs which can be drawn without intersecting edges) is of particular interest for our application since it is often the case for water supply networks. Finding the optimal solution is NP-hard also for the planar case, however the *planar separator theorem* (Lipton and Tarjan 1979) states that a bisection in which the biggest set contains at most two thirds of the vertices and whose separator contains $O(\sqrt{n})$ vertices can be found in linear time.

Other related problems are the multicut problems (Pichler et al. 2010), in which the aim is to find the minimal set of edges (or nodes) such that given pairs of nodes are no longer connected. In our case, instead, the aim is to disconnect a possibly small part of the network while keeping connected all the rest.

The algorithms for graph partitioning or solving multicut problems are clearly not directly applicable to the valve placement problem, also because of the issue of unintended isolation mentioned in Section 1. However, it would be interesting to hybridise our algorithm with some of the techniques available for such problems; we plan to study the feasibility of such approaches in future work.

## 8 Conclusions and future work

We presented a new application, taken from the hydraulic domain, for logic programming. We proposed an algorithm based on CLP(FD), and found solutions better than the best solutions known in the literature. The computation time can be high when the number of valves is high, but in many cases it is still acceptable since it is applied during the design of a water distribution network.

The model can be improved in various directions. Taking cue from the findings of Creaco et al. (2010), the variable "weighted average demand shortfall", based on the likelihood (or, more in general, on the probability relative to a prefixed time interval) of failures occurring within the segments as a result of mechanical factors could be used. In fact it has been observed that this variable is superior to the variable "maximum demand shortfall". Weighted average demand shortfall takes into account the entire network and not only the largest segment in terms of demand shortfall; moreover, it guarantees that unique solutions will be found for a pre-established number or cost of the valves in the network.

The current implementation runs on an open source Prolog, but very fast commercial ones, such as SICStus or B-Prolog (Zhou 2011) may provide a strong speedup; in the future we plan to port the implementation on other Prolog systems. In fact, there exist a library for graph algorithms also in SICStus, while in B-Prolog the table mode (Warren 1992; Zhou et al. 2008) lets one easily implement efficient graph predicates. Unluckily, the syntax of ECL$^i$PS$^e$ and SICStus/B-Prolog is different, although similar; for example, although both ECL$^i$PS$^e$ (Schimpf 2002) and B-Prolog (Zhou 2011) support logical loops, they adopt a different syntax.

From the computational viewpoint, other techniques could be adopted in order to reduce the computation time. For example, we could apply restarts (Gomes et al. 1997), or try better heuristics to select the next variable to be assigned; from general ones, like, for example, *dom/wdeg* (Boussemart et al. 2004), to specific ones.

When the number of valves is very high, we could directly apply incomplete methods, that try to get quickly good solutions sacrificing the proof of optimality. One very promising approach would be to use Large Neighbourhood Search, that has been implemented in Prolog in previous works (Dal Palù et al. 2010),

A very interesting line of research would be to develop an Answer Set Program-ming model. ASP is often faster than CLP(FD) when the domains of the variables do not contain many values (Dovier et al. 2005; Mancini et al. 2008); we plan to develop an ASP application in the near future.